%% file: ltfatnote050.tex
\documentclass[conference]{IEEEtran}

\ifCLASSINFOpdf
   \usepackage[pdftex]{graphicx}
   \graphicspath{{../pdf/}{../jpeg/}}
   \DeclareGraphicsExtensions{.pdf,.jpeg,.png}
\else
   \usepackage[dvips]{graphicx}
   \graphicspath{{../eps/}}
   \DeclareGraphicsExtensions{.eps}
\fi

\usepackage{url}

\usepackage{amsmath,amssymb,amsfonts,amsthm}
\usepackage[latin1]{inputenc}
\usepackage[T1]{fontenc}
\usepackage{ifpdf}
\usepackage[english]{babel}
\usepackage[labelfont=normalsize,textfont=normalsize]{subfig} 

 \usepackage{algorithm2e}

\usepackage{tikz}
\usetikzlibrary{arrows}

\newcommand{\gana}{g_\text{a}}

\newcommand{\aana}{a_\text{a}}
\newcommand{\asyn}{a_\text{s}}
\newcommand{\bana}{b_\text{a}}
\newcommand{\bsyn}{b_\text{s}}
\newcommand{\phaseana}{\phi_\text{a}}
\newcommand{\phasesyn}{\phi_\text{s}}

\newcommand{\tgrad}{\left( \Delta_\text{t}\phi_\text{a}\right)}
\newcommand{\fgrad}{\left( \Delta_\text{f}\phi_\text{a}\right)}
\newcommand{\me}{\mathrm{e}}
\newcommand{\mi}{\mathrm{i}}
\newcommand{\princarg}[1]{\left\lbrack #1 \right\rbrack_{2\pi}}

\begin{document}
\title{Phase Vocoder Done Right}

\author{
\IEEEauthorblockN{Zden\v{e}k Pr\r{u}\v{s}a and Nicki Holighaus}
\IEEEauthorblockA{%
Acoustics Research Institute, Austrian Academy of Sciences\\
Vienna, Austria\\
\{zdenek.prusa,nicki.holighaus\}@oeaw.ac.at
}}

\maketitle

\begin{abstract}
    The phase vocoder (PV) is a widely spread technique for
    processing audio signals. It employs a short-time Fourier transform
    (STFT) analysis-modify-synthesis loop and is typically used for 
    time-scaling of signals by means of 
    using different time steps for STFT analysis and synthesis.
    The main challenge of PV used for that purpose
    is the correction of the STFT phase.
    In this paper, we introduce a novel method for phase correction 
    based on phase gradient estimation and its integration. 
    The method does not require explicit peak picking and tracking nor does
    it require detection of transients and their separate treatment.
    Yet, the method does not suffer from the typical phase vocoder artifacts
    even for extreme time stretching factors.
    %
\end{abstract}

\IEEEpeerreviewmaketitle

\section{Introduction}
\label{sec:intro}
The term \emph{phase vocoder} was coined by Flanagan and Golden
\cite{flgo66}
but the now classical form of PV employing STFT analysis and synthesis
using different analysis and synthesis steps
for the purposes of time-scaling was introduced later 
by Portnoff~\cite[Sec. 6.4.4]{po78}.
The way how the phase is corrected in the classical
PV is based on the linear phase progression of a sinusoid with
constant frequency. Each frequency channel is treated as a 
separate sinusoid whose phase is computed by accumulating 
the estimate of its instantaneous frequency and thus preserving 
the \emph{horizontal phase coherence} \cite{mola95}.
Obviously, such an approach cannot cope well with
time-varying sinusoidal signals and non-sinusoidal signals.
A modification of PV was proposed by Laroche and Dolson \cite{lado97,lado99}.
This improved PV involves spectral peak picking, tracking and ``locking'' the
phase of the frequency bins belonging to a sinusoid to the phase of the peak
(\emph{scaled phase locking}). The phase locking enforces 
\emph{vertical phase coherence} within the region of influence of the peak.
Although the phase-locked PV is able to handle signals consisting of sinusoids with time-varying frequencies, it still fails for percussive sounds and transient events in general.
A review of the phase vocoder, its history and alternative
approaches to time and frequency scale modifications of audio signals
can be found in \cite{zo02,liro13,drmu16}.

Time scaling using phase vocoder techniques is known to produce specific artifacts
such as transient smearing, ``echo'' and ``loss of presence''
collectively referred to as \emph{phasiness} \cite{lado97}.
The artifacts are generally attributed to the loss of vertical phase coherence.
Transient smearing compensation has been addressed by several authors. 
The most common approach is performing a \emph{phase reset} or
phase locking at transients
\cite{dudasa02,ro03,rasabe05}. Other approaches involve disabling
the time-scaling at transients
\cite{nawa09}, or rely on using different window lengths for harmonic and transient parts
\cite{drmuew14, otdo16}. 
All approaches mentioned rely on correct transient detection. 
The preservation of vertical phase coherence is considered to be important for the
quality of time-scaled voiced speech signals. It has been reported that not
preserving the relative phase shift between the fundamental and the partials
is perceived as unnatural.
Several specialized methods for time-stretching of monophonic voice signals
were proposed \cite{ro10,modu11}.
Historically, time scaling techniques which preserve
relative phase shift between the partials
are called \emph{shape preserving} \cite{qumc92}.
To avoid dealing with phase 
altogether, a magnitude-only reconstruction has been considered \cite{griflim84}.
Although efficient and real-time algorithms have recently been proposed, e.g. \cite{ltfatnote043,ltfatnote048}, 
they do not perform favorably for large stretching factors. This is due to the magnitude
not complying with the new synthesis step. 

In this paper, we propose a novel method for phase correction
in the PV, relying on both partial derivatives of the STFT phase and {their integration.}
The phase derivatives are estimated using centered finite differences
and the integration is performed through real-time phase gradient heap integration algorithm (RTPGHI)
\cite{ltfatnote043}, an extension of the PGHI algorithm \cite{ltfatnote040}, 
originally proposed for magnitude-only reconstruction.
The algorithm automatically enforces horizontal and vertical phase
coherence even for broadband, non-sinusoidal components. No explicit peak 
picking or transient detection is required.

\section{Classical PV Revisited}
\label{sec:clasPV}
In this section, we recall the essential formulas connected to PV
and explain the role of the partial STFT phase derivative in the time direction.
We further show with a straightforward example, that the poor performance of 
the classical PV for non-sinusoidal components can be attributed to 
neglecting the frequency direction partial phase derivative.

Given a discrete time signal $f$ which is nonzero on the interval $0,\dots,L-1$,
a real-valued analysis window $\gana$ concentrated around the origin
and the analysis time step $\aana$, 
the discrete STFT is given by
\begin{equation}
    c(m,n) = \sum_{l\in\mathbb{Z}}
    f(l+n\aana)
        \gana(l)
    \me^{-\mi 2 \pi m l / M} \label{eq:stft}
\end{equation}
for $m=0,\dots,M-1$, $M$ being the FFT length, and $n=0,\dots,N -1$, where $N=L/\aana$
is the number of STFT frames. 
Setting $M=L$ results in the full frequency resolution, but,
typically, it is chosen such that $M\ll L$.
We define the analysis frequency step as
\begin{equation}
    \bana = \frac{L}{M}.
\end{equation}
The magnitude $s$ and phase $\phi_\text{a}$ components of the STFT can be separated by
\begin{align}
    s(m,n) &= \left| c(m,n) \right|,  \\ 
    \phi_\text{a}(m,n)  &= \arg \left( c(m,n) \right).
\end{align}
The time scaling factor will be defined as
\begin{equation}
    \alpha = \frac{\asyn}{\aana},
    \label{eq:scaling1}
\end{equation}
where $\asyn$ denotes the synthesis time step. 
The length of the output signal is therefore
equal to $\alpha L$ and the 
synthesis frequency step to
\begin{equation}
    \bsyn = \alpha{\bana}.
    \label{eq:scaling2}
\end{equation}
The synthesis phase is constructed by the recursive 
\emph{phase propagation } (integration) formula \cite{lado99}
\begin{equation}
        \phi_\text{s}(m,n) =  \phi_\text{s}(m,n-1) +\asyn \tgrad(m,n),
    \label{eq:synphase1}
\end{equation}
where $\Delta_\text{t}$ performs differentiation of the analysis phase $\phi_\text{a}$
(see Section \ref{sec:pgi} for details),
or alternatively, by employing the trapezoidal integration rule, by
\begin{equation}
    \begin{split}
        \phi_\text{s}(m,n) &=  \phi_\text{s}(m,n-1) + {}\\
        &+\frac{\asyn}{2}\left(  \tgrad(m,n-1)
    + \tgrad(m,n)\right).
    \end{split}
    \label{eq:synphase2}
\end{equation}
The time-scaled signal $\widetilde{f}$ is reconstructed using
\begin{align}
    \widetilde{f}(l) &= 
    \sum_{n=0}^{N-1} \widetilde{f_n}(l-n\asyn) 
    \ \ \ \text{for} \ l=0,\dots,\alpha L-1 \ \text{with}\\
    \widetilde{f_n}(l) &=
    g_\text{s}(l) 
    \sum\limits_{m=0}^{M-1} s(m,n)\me^{\mi\phi_\text{s}(m,n)}\me^{\mi 2 \pi m l / M}
    \label{eq:istft}
    \ \ \text{and with}\\
    g_\text{s}(l) &=
    \frac{1}{M}\frac{g_\text{a}(l)}{\sum_{n\in\mathbb{Z}}g_\text{a}(l - n\asyn)^2}
\end{align}
being the synthesis window $g_\text{s}$.
To summarize, the classical PV performs differentiation and back integration of
the phase in the time direction for all frequency bins.

Formula \eqref{eq:stft} can be interpreted as a sampling of the
continuous STFT which is by itself a two-dimensional function of time and frequency.
Therefore, $\tgrad$ can be interpreted as an approximation of the 
partial derivative of the phase in time.
The complete phase gradient however involves also the partial derivative in 
frequency whose approximation will further be denoted as $\fgrad$. 
This second gradient component is completely disregarded in the classical PV,
which introduces significant inaccuracies except for pure, stationary sinusoids.
For the latter, $\fgrad$ indeed equals zero.
The example in Figures~1A--1C shows plots of a spectrogram of
a sum of a pure sinusoid, an exponential chirp and an impulse 
and of the partial derivatives of the phase. In particular, Figure 1C shows that,
for the chirp and pulse components, $\fgrad$ unsurprisingly has non-negligible values.

\begin{figure}[!th]
    \centering
    \subfloat[Spectrogram of a sum of a pure sinusoid,
    an exponential chirp and an impulse. The values are in dB.]{%
        \includegraphics[trim={0 0 0 0},clip,width=0.99\linewidth]{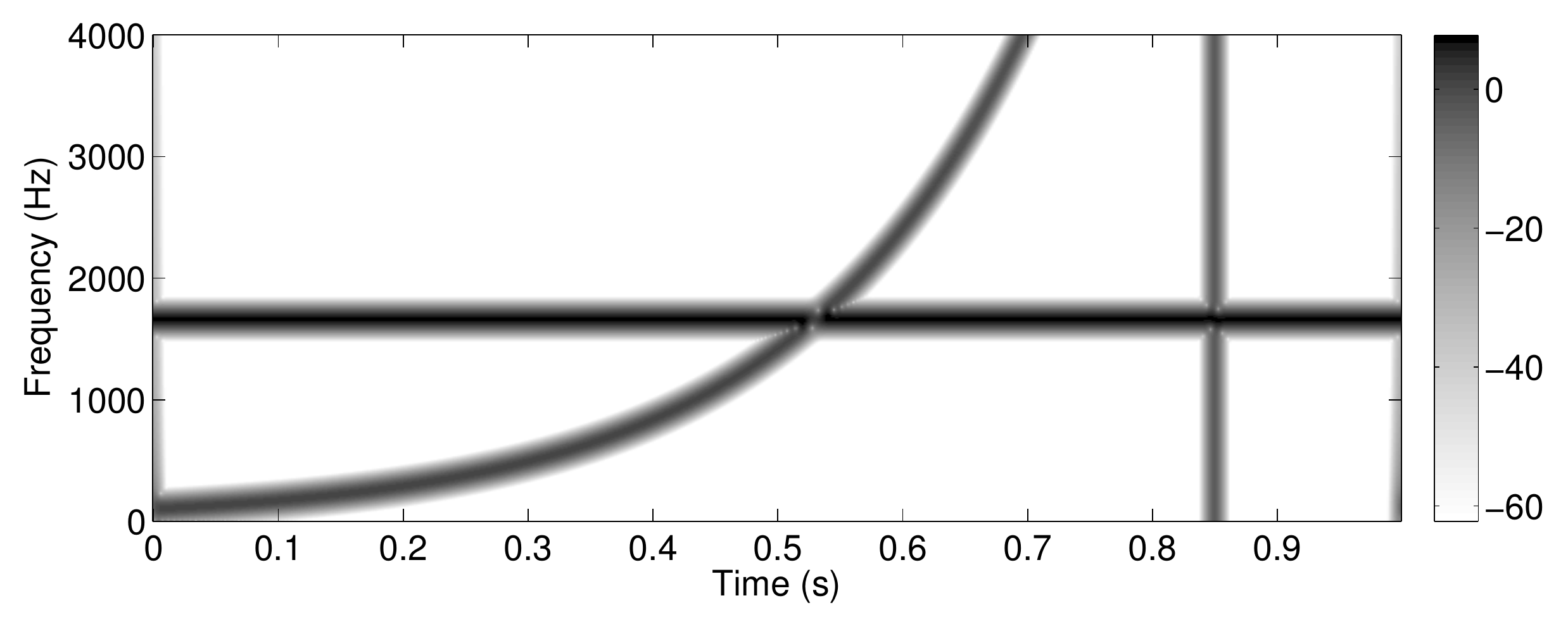}
        \label{subfig:sinchirp}
    }\vspace{-1em}\\%
    \subfloat[][%
        Scaled time phase derivative $\frac{f_\text{s}}{2\pi}\tgrad$ (instantaneous
        frequency). $f_\text{s}$ stands for the sampling rate and the values are in Hertz.
    ]{%
        \includegraphics[trim={0 0 0 0},clip,width=0.99\linewidth]{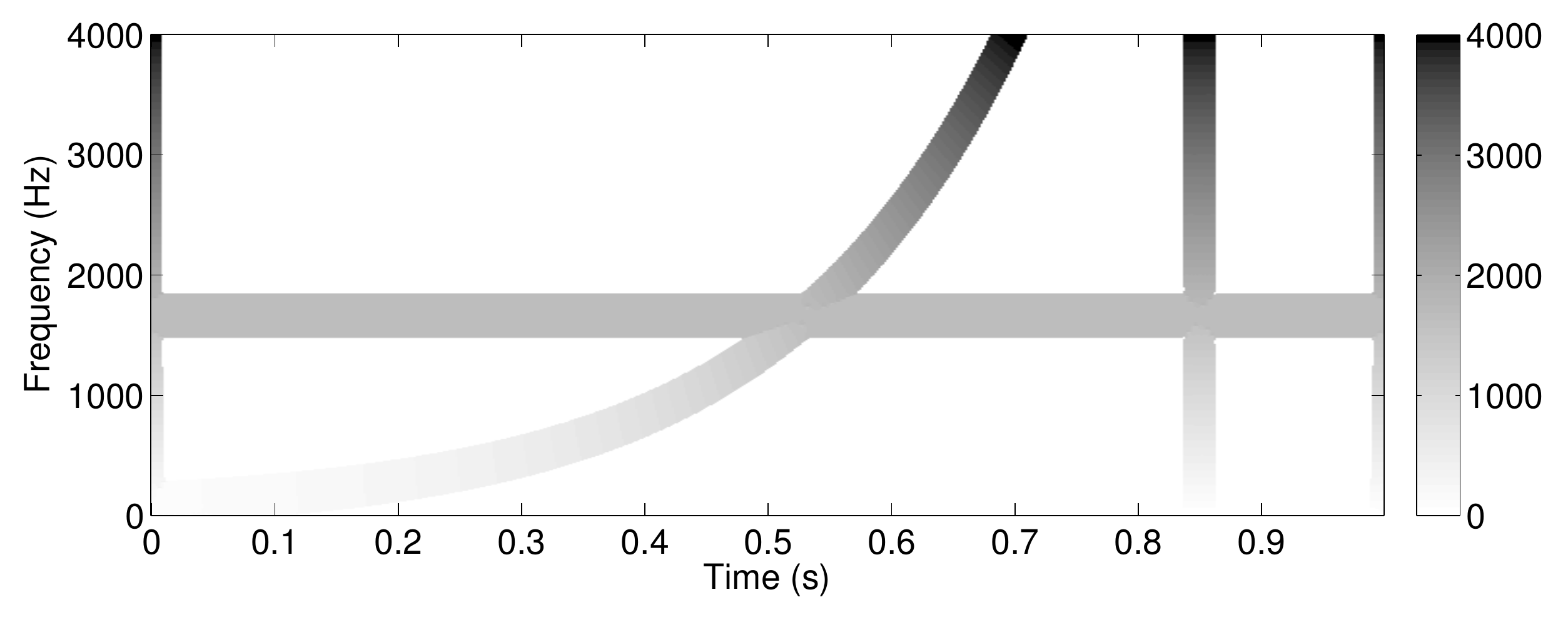}
        \label{subfig:sinchirp_tgrad}
    }\vspace{-1em}\\ 
    \subfloat[][%
        Scaled abs. value of the frequency phase derivative $\frac{10^3L}{2\pi f_\text{s}}\left| \fgrad
        \right|$ (abs. val. of the local group delay). Note that $\fgrad$ is zero for the pure sinusoid. The values are in milliseconds.
        ]{%
        \includegraphics[trim={0 0 0 0},clip,width=0.99\linewidth]{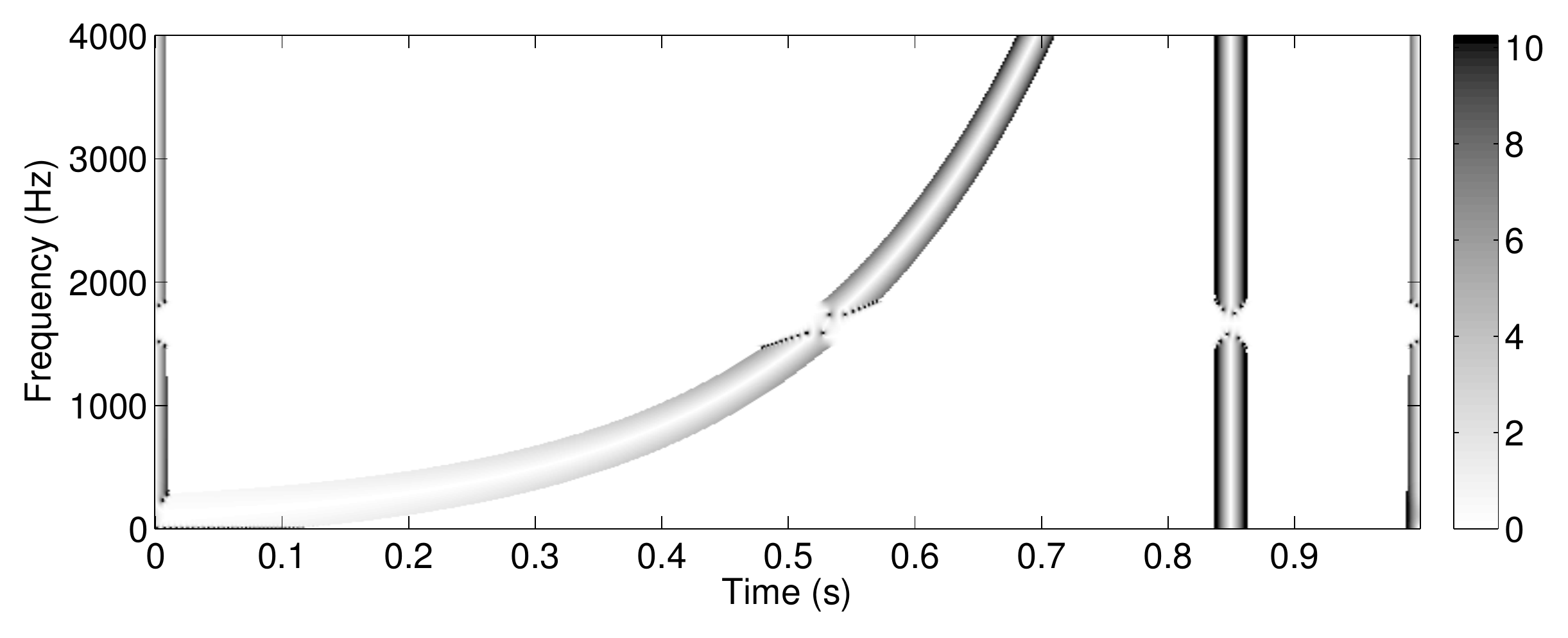} \
        \label{subfig:sinchirp_fgrad}
    }    
    \phantomcaption\label{fig:phasegrad}    
\end{figure}


\section{Full Phase Gradient Estimation And Integration}
\label{sec:pgi}
In this section, we present formulas for estimating 
partial derivatives of the STFT phase $\fgrad$ and $\tgrad$ and present an
adaptive integration algorithm which takes them both into account to produce
an adjusted synthesis phase $\phasesyn$ to be used in \eqref{eq:istft}.

The formulas exploit the well known phase \emph{unwrapping} procedure
which involves taking the principal argument of an angle.
We adopt the notation from \cite{krhoknzo12} and define the principal 
argument as
\begin{equation}
    \princarg{x} = x - 2\pi\left\lceil \frac{x}{2\pi} \right\rfloor, 
    \label{eq:princarg}
\end{equation}
where $\lceil \cdot \rfloor$ denotes rounding towards the closest integer.

The well known phase differentiation in the time direction
(involving conversion to \emph{heterodyned} \cite{lado99} phase difference and back)
can be written as 
\begin{equation}
    \begin{aligned}
    & \left( \Delta_\text{t,back} \phaseana \right)(m,n)  =  \\
        &\frac{1}{\aana}
        \princarg{%
        \phaseana(m,n) - \phaseana(m,n-1) - \frac{2\pi m\aana}{M}
        }
        + \frac{2\pi m}{M},
    \label{eq:backdiff}
    \end{aligned}
\end{equation}
which is a modified backward difference scheme. The forward
difference scheme can be written similarly
\begin{equation}
    \begin{aligned}
    & \left( \Delta_\text{t,fwd} \phaseana \right)(m,n)  =  \\
        &\frac{1}{\aana}
        \princarg{%
        \phaseana(m,n+1) - \phaseana(m,n) - \frac{2\pi m\aana}{M} 
        }
        + \frac{2\pi m}{M}
    \label{eq:fwddiff}
    \end{aligned}
\end{equation}
and the centered difference is simply an average of the two
\begin{equation}
    \begin{aligned}
        &\left( \Delta_\text{t,cent} \phi_\text{a} \right)(m,n)
        = \\ 
        &\frac{1}{2}
        \left( 
        \left(\Delta_\text{t,back} \phaseana \right)(m,n) +
        \left(\Delta_\text{t,fwd} \phaseana \right)(m,n)
        \right).
    \label{eq:centdiff}
    \end{aligned}
\end{equation}
Any of the schemes can be used in place of $\Delta_\text{t}$.
In our experience $ \Delta_\text{t,cent}$ is the most suitable.
Similarly, the differentiation in the frequency direction 
can be performed using the backward scheme
\begin{equation}
    \left( \Delta_\text{f,back} \phaseana \right)(m,n) = 
    \frac{1}{\bana} \princarg{ \phaseana(m,n) - \phaseana(m-1,n) },
    \label{eq:fgradback}
\end{equation}
the forward scheme
\begin{equation}
    \left( \Delta_\text{f,fwd} \phi_\text{a} \right)(m,n) = 
    \frac{1}{\bana} \princarg{ \phaseana(m+1,n) - \phaseana(m,n) },
    \label{eq:fgradfwd}
\end{equation}
or the centered scheme
\begin{equation}
    \begin{aligned}
        &\left( \Delta_\text{f,cent} \phaseana \right)(m,n)
        = \\
        &\frac{1}{2}
        \left( 
        \left(\Delta_\text{f,back} \phi_\text{a} \right)(m,n) +
        \left(\Delta_\text{f,fwd} \phi_\text{a} \right)(m,n)
        \right).
    \label{eq:fgradcent}
    \end{aligned}
\end{equation}
%

Having the means for estimating the phase 
partial derivatives, 
the RTPGHI algorithm \cite{ltfatnote043}
can be invoked with few modifications.
In its essence, the method proceeds by processing one frame at a time
computing the synthesis phase of the current $n$-th frame $\phasesyn(\cdot,n)$.
It requires storing the already computed phase $\phasesyn(\cdot,n-1)$
and the time derivative $\tgrad(\cdot,n-1)$
of the previous $(n-1)$-th frame and further, it requires access to
the coefficients of the previous, current and one ``future''
frame ($c(\cdot,n-1)$, $c(\cdot,n)$ and $c(\cdot,n+1)$)
assuming the centered differentiation scheme \eqref{eq:centdiff} is used for computing
$\tgrad(\cdot,n)$.
Before the algorithm starts $\tgrad(m,n)$ and $\fgrad(m,n)$ are computed for all
$m$ and current $n$.
The algorithm starts by selecting the frequency bin $m_{\text{h}}$ of the
coefficient with the highest magnitude from the previous $(n-1)$-th frame and 
propagates the phase along the time direction to the current frame using 
$\eqref{eq:synphase2}$ such that $\phasesyn(m_{\text{h}},n)$
is obtained. Up to this point, the procedure is equivalent with the classical PV.
The way how the phase of the remaining coefficients is obtained is different
since the phase can now be propagated also in the frequency direction from
the already computed phase in the current frame.
The phase propagation direction is decided on-the-fly according to the
magnitude of the coefficients.

The steps of the algorithm are summarized in Alg.~\ref{alg:rtpghi}.
\begin{algorithm}[!ht]
    \LinesNumbered
    \caption{Phase Gradient Heap Integration for $n$-th frame}
    \label{alg:rtpghi}
    \KwIn{Phase time derivative $\tgrad$
        and magnitude $s$ of frames $n$ and $n-1$,
        phase frequency derivative $\fgrad$ for frame $n$,
        estimated phase $\phasesyn$ for frame $n-1$
    and relative tolerance $\mathit{tol}$.}
    \KwOut{Phase estimate $\phasesyn$ for frame $n$.}
    $\mathit{abstol} \leftarrow \mathit{tol}\cdot \max\left( s(m,n) \cup s(m,n-1)  \right)$\;
    Create set $\mathcal{I}=\left\{m : s(m,n)>\mathit{abstol}
    \right\}$\;\label{alg:setI}
    Assign random values to $\phasesyn(m,n)$ for $m\notin \mathcal{I}$\;
    Construct a self-sorting max \emph{heap} \cite{wi64} for $(m,n)$ tuples\;\label{alg:line}
    Insert $(m,n-1)$ for $m\in \mathcal{I}$ 
    into the \emph{heap}\;\label{alg:mark}
    \While{$\mathcal{I}$ is not $\emptyset$  }{
            $(m_{\text{h}},n_{\text{h}}) \leftarrow$ remove the top of the
            \emph{heap}\;\label{alg:heapremove}

            \If{$n_{\text{h}}=n-1$}{
                \If{$(m_{\text{h}},n) \in \mathcal{I}$}{
                    $\phasesyn(m_{\text{h}},n) \leftarrow \phasesyn(m_{\text{h}},n-1) +
                    \frac{\asyn}{2}\left( \tgrad(m_{\text{h}},n-1) +
                    \tgrad(m_{\text{h}},n)  \right)$\;\label{alg:initpoint}
                    Remove $(m_{\text{h}},n)$ from  $\mathcal{I}$\;\label{alg:initfromI}
                    Insert $(m_{\text{h}},n)$ into the \emph{heap}\;\label{alg:inittoheap}
                }
            }
            \If{$n_{\text{h}}=n$}{\label{alg:nframe}
                \If{$(m_{\text{h}}+1,n) \in \mathcal{I}$}{
                    $\phasesyn(m_{\text{h}}+1,n) \leftarrow \phasesyn(m_{\text{h}},n) +
                    \frac{\bsyn}{2}\left( \fgrad(m_{\text{h}},n) + \fgrad(m_{\text{h}}+1,n)  \right)$\;\label{alg:above}
                    Remove $(m_{\text{h}}+1,n)$ from  $\mathcal{I}$\;
                    Insert $(m_{\text{h}}+1,n)$ into the \emph{heap}\;
                }
                \If{$(m_{\text{h}}-1,n) \in \mathcal{I}$}{
                    $\phasesyn(m_{\text{h}}-1,n) \leftarrow \phasesyn(m_{\text{h}},n) -
                    \frac{\bsyn}{2}\left( \fgrad(m_{\text{h}},n) + \fgrad(m_{\text{h}}-1,n)  \right)$\;\label{alg:below}
                    Remove $(m_{\text{h}}-1,n)$ from  $\mathcal{I}$\;
                    Insert $(m_{\text{h}}-1,n)$ into the \emph{heap}\;
                }
        }
    }
    \bigskip
\end{algorithm}
Note that the algorithm employs trapezoidal integration rule for the phase
propagation on lines
\ref{alg:initpoint}, \ref{alg:above} and \ref{alg:below}. Line \ref{alg:initpoint}
implements \eqref{eq:synphase2}, but the integration is also performed in
the vertical (frequency) direction (lines \ref{alg:above}, \ref{alg:below})
employing the frequency phase derivative estimate $\fgrad$.
The phase for each frequency bin is however computed only once.
Which equation is used is decided adaptively according to the magnitude $s$ with
the help of the \emph{max heap} \cite{wi64}. 
The max heap is a dynamic data structure that always 
places coordinates $(m,n)$ of the coefficient with the highest magnitude on top.
It is populated by the coordinates of the coefficients with already known phase which are 
possible candidates from which the phase can be propagated further.
At the beginning of the algorithm, all frequency bins from the
previous frame with significant magnitude are inserted into the heap (line 5).
In the first iteration, the only possible phase propagation direction
is the time direction on line \ref{alg:initpoint}. The coordinates of the
just computed coefficient are inserted into the heap (line 13).
In the next iteration, two options are possible. Either
the same procedure is repeated with another frequency bin from the previous frame,
or the extracted top of the heap belongs to the current frame $n$.
In that case, the phase is propagated in the frequency direction
to both neighboring frequency bins on lines \ref{alg:above} and \ref{alg:below}.
Both neighbors are then inserted into the heap and in the next step,
the integration continues with extracting the new heap top and spreading the 
phase further.

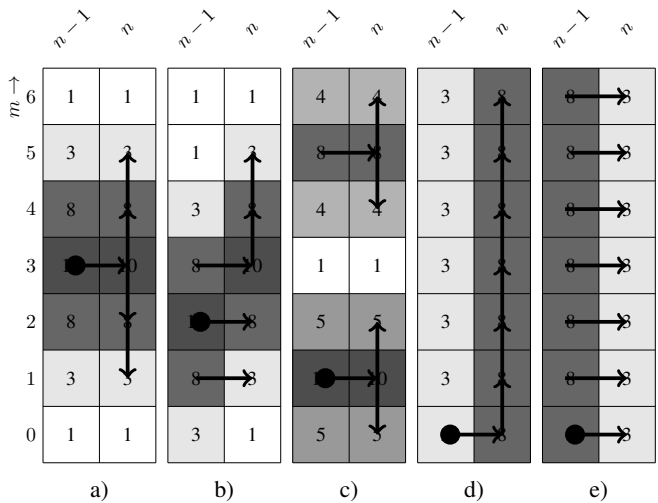
\begin{figure}[!t]
    \input{rtpghiexamples.tex}
    \caption{Conceptual spectrograms overlaid with the phase spreading paths: %
    a) pure sinusoid, b) linear chirp, c) two sinusoids, d) before impulse peak, 
    e) after impulse peak.
    Coefficients with magnitude 1 are below the tolerance of the
    algorithm.
    }
    \label{fig:rtpghiexamples}
\end{figure}

Fig.~\ref{fig:rtpghiexamples} shows examples of how the phase is spread for
several signals. 
The orientation of the arrow identifies the line in the algorithm as follows: 
\ref{alg:initpoint} ($\rightarrow$), \ref{alg:above} ($\uparrow$) and
\ref{alg:below}($\downarrow$).

\section{Evaluation}
\label{sec:evaluation}
This section is devoted to the testing and evaluation of the proposed PV.
We present the STFT setting used and summarize results of a listening test. 

In the evaluation of the proposed PV we used a 4092 samples long Hann window, 
set the number of frequency channels to $M=8192$ and fixed the
synthesis time step to $\asyn=1024$, changing $\aana$ according to
\eqref{eq:scaling1} rounded to the closest integer. 
All sound excerpts were sampled at 44.1~kHz.
The tolerance in Algorithm~\ref{alg:rtpghi} was set to $\mathit{tol}=10^{-6}$.
We compared the results with the following algorithms and software:
\begin{itemize}
    \item Classical PV (\textbf{PV}) using the same setting as the proposed algorithm.
    \item \'{e}lastique PRO (\textbf{EL}) \cite{elastique} 
(accessed through the Sonic API \url{http://www.sonicapi.com} with the help of a
script from the TSM toolbox \cite{drmu14}). 
    \item IRCAM Lab TS 1.0.11 (\textbf{IR}) \cite{ircamts} (demo version)
    with the quality set to high, 
    polyphonic setting and enabled transient preservation.
\item melodyne4 4.1.0 (\textbf{ME}) \cite{melodyne} (trial version) with the universal algorithm.
\end{itemize}
It is worth mentioning that both \textbf{IR} and \textbf{ME} offer additional
modes specialized to specific types of audio signals (voice, percussions, etc.). 
We have intentionally used the most universal options.

The listening tests were performed for $\alpha= \left\{1.5, 2\right\}$ using 
a web interface \cite{waet2015}. 
In the tests, we used the proposed algorithm as the reference and asked
subjects to rate presented sound examples on the relative  
comparison category rating (CCR) scale \cite{itute800}
({\it 3 Much better, 2 Better, 1 Slightly better, 0 About the same,
 $-1$ Slightly worse, $-2$ Worse, $-3$ Much worse}).
Each time, 5 sound examples were rated;
4 for the algorithms and software mentioned above and 1
was a hidden reference in the fashion of the MUSHRA test \cite{itute1534} 
(similar tests were performed in \cite{krhoknzo12}).
A total number of 8 subjects with background in acoustics participated in the test.
Two subjects were removed from the evaluation because they
regularly failed to identify the hidden reference.
Table~\ref{tab:excerpts} is a list of excerpts used and
Table~\ref{tab:results} shows the average CCR for the individual
sound excerpts. 
As expected, the classical PV (\textbf{PV}) was consistently rated worse
than the proposed method.

For $\alpha=1.5$, \textbf{EL} and \textbf{IR} were rated 
close to the proposed method, with the exception of
Latino and DrumSolo for \textbf{EL} which were slightly worse
mainly due to poorer transient preservation.
\textbf{ME} was rated slightly worse consistently.

For $\alpha=2$, both \textbf{EL} and \textbf{ME} were rated worse than
the proposed method consistently.
For \textbf{IR}, complex mixes PeterGabriel and EddieRabbit were rated slightly better
than the proposed algorithm. 
We suspect that this is mainly due to the active transient preservation in \textbf{IR},
which results in a shorter attack time of the transients.
On the other hand, 
the female voice in Musetta is slightly more ``phasy''
than the proposed method, which could be a reason for why it was 
rated slightly worse.
%
%
%

All sound examples are available at \url{http://ltfat.github.io/notes/050}.

\begin{table}[!t]
    \centering
    \caption{List of sound examples}
    \label{tab:excerpts}
    \begin{tabular}{l|p{0.7\linewidth}}
        \textbf{Name}       & \textbf{Description} \\ \hline\hline
        CastViolin & Solo violin overlaid with castanets from
        \cite{drmu14}.\\ \hline
        DrumSolo        & A solo on a drum set from \cite{drmu14}.\\ \hline
        Latino          & Pan flute, guitar and percussion.\\ \hline
        Musetta         & An excerpt from a song \emph{Ophelia's song} performed by
                          \emph{Musetta} from the \emph{Nice to meet you!} album
                          (female solo singing).
                          \\ \hline
        March   & An excerpt from \emph{Radetzky March} by \emph{Johann Strauss}
                        performed by an orchestra.\\ \hline
        PeterGabriel    & An excerpt from a song \emph{Sledgehammer} performed by
        \emph{Peter Gabriel} from the album \emph{So}. \\ \hline
        EddieRabbit     & An excerpt from a song \emph{Early in the morning} performed by
                         \emph{Eddie Rabbit} from the album \emph{Step by Step}.\\
    \end{tabular}
\end{table}
{\setlength{\tabcolsep}{0.33em}
\begin{table}[!t]
    \centering
    \caption{Comparison mean opinion score for sound examples}
    \label{tab:results}
    \begin{tabular}{l| r r r r | r r r r }
        & \multicolumn{4}{c|}{$\alpha=2$} & \multicolumn{4}{c}{$\alpha=1.5$} \\
        & \textbf{PV} & \textbf{EL} & \textbf{ME} & \textbf{IR} & \textbf{PV} &
        \textbf{EL} & \textbf{ME} & \textbf{IR} \\ \hline \hline
        CastViolin      &  $-2.7$ & $-0.4$  & $-0.8$& $-0.3$    
                        &  $-2.3$ & $0.3$   & $-1.0$& $-0.3$ \\ 
        DrumSolo        &  $-2.9$ & $-2.0$  & $-2.7$& $0.5$     
                        & $-2.8$  & $-0.8$  & $-0.6$& $0.0$\\
        Latino          &  $-1.4$ & $-1.3$  & $-0.3$& $0.3$  
                        &  $-1.7$ & $-0.9$  & $-1.5$& $-0.5$ \\
        Musetta         &  $-1.6$ & $-1.0$  & $-2.1$& $-0.8$  
                        & $-2.1$  & $0.2$   & $-0.8$& $0.1$\\
        March           &  $-1.3$ & $-1.8$  & $-1.1$& $-0.7$    
                        & $-1.3$  & $-0.6$  & $-0.7$& $-0.3$ \\
        PeterGabriel    &  $-1.1$ & $-1.8$  & $-1.2$& $0.9$ 
                        & $-2.0$  & $-0.5$  & $-1.1$& $-0.2$\\
        EddieRabbit     &  $-0.7$ & $-2.0$  & $-1.4$ & $1.3$ 
                        & $-2.2$  & $-0.2$  & $-0.8$& $0.1$ \\ \hline
        Avg.            &  $-1.7$ & $-1.5$  & $-1.4$& $0.2$  
                        & $-2.1$  & $-0.3$  & $-0.9$& $-0.2$
    \end{tabular}
\end{table}
}

\section{Conclusion}
We presented a novel method for adjusting phase in 
PV used for time-scaling of audio signals.
The method is simple in the sense that it works automatically and
it does not require analyzing the signal content.
The listening test shows that it is competitive with 
state-of-the-art time-stretching in commercial software solutions.

While the proposed method does preserve transients without the
necessity of detecting them, it also stretches them together 
with the rest of the signal.
If true transient preservation, or rather \emph{transient sharpening}
is desired, their detection
and special treatment seems to be unavoidable.
The transient processing method by R\"obel \cite{ro03}
is a suitable candidate for such task as it uses already
computed $\fgrad$ for transient detection.

Unfortunately, the current implementation of the proposed method 
is not optimal for voiced monophonic speech signals. We expect that 
this is due to the possible phase shift between the fundamental and the partials.
Specialized PV variants (e.g. Moinet and Dutoid \cite{modu11}) provide perceptually 
better results.
Therefore, in the future, we plan to enhance the proposed PV
with the shape-preserving property. Moreover, an extension of PGHI to more general 
filter banks has been proposed in \cite{ltfatnote051}. The design 
of a phase vocoder employing a filter bank with variable time-frequency 
resolution could provide an interesting starting point for further 
developments.

\section*{Acknowledgment}
The authors thank all participants of the listening test
and Thibaud Necciari for valuable comments and suggestions.

This work was supported by the Austrian Science Fund (FWF): Y~551--N13 and I~3067--N30.

\bibliographystyle{IEEEtran}
\bibliography{project} 

\end{document}

%% file: rtpghiexamples.tex
\definecolor{color10}{gray}{0.3}
\definecolor{color9}{rgb}{0.9731,   0.8054,    0.2169}
\definecolor{color8}{gray}{0.4}
\definecolor{color7}{rgb}{0.9265,    0.4054,    0.1503}
\definecolor{color6}{rgb}{0.8071 ,   0.2627,    0.2789}
\definecolor{color5}{gray}{ 0.6 }
\definecolor{color4}{gray}{0.7}
\definecolor{color3}{gray}{ 0.9}
\definecolor{color1}{gray}{1}
\definecolor{color0}{gray}{0.5}
\colorlet{othergray}{blue}

\newcommand*{\xMin}{0}%
\newcommand*{\xMax}{2}%
\newcommand*{\xMaxMin}{1}%
\newcommand*{\yMin}{0}%
\newcommand*{\yMax}{7}%
\newcommand*{\yMaxMin}{6}%
\newcommand{\myscale}{0.75}

\newcommand{\mysquare}[3]{\fill[color#3] (#1,#2) rectangle ({#1+1},{#2+1}) node [black]
at ({#1+0.5},{#2+0.5}) {#3};}

\newcommand{\mysquaree}[2]{\fill[color0] (#1,#2) rectangle ({#1+1},{#2+1});}
\newcommand{\mysquareee}[3]{\fill[#3] (#1,#2) rectangle ({#1+1},{#2+1});}

\newcommand{\mygrid}{%
    \foreach \i in {\xMin,...,\xMaxMin} {
        \draw [very thin,black] (\i,\yMin) -- (\i,\yMax) ;
    }

    \draw [very thin,black] (\xMax,\yMin) -- (\xMax,\yMax);

    \foreach \i in {\yMin,...,\yMaxMin} {
        \draw [very thin,black] (\xMin,\i) -- (\xMax,\i) ;
    }

    \draw [very thin,black] (\xMin,\yMax) -- (\xMax,\yMax);

    \draw [black,rotate around={45:(0.5,7.7)}] node at (0.5,7.7) {$n-1$};
    \draw [black,rotate around={45:(1.5,7.7)}] node at (1.5,7.7) {$n$};
}

    \begin{tikzpicture}[scale=\myscale, every node/.style={transform shape}]

        \mysquare{0}{0}{1}
        \mysquare{0}{1}{3}
        \mysquare{0}{2}{8}
        \mysquare{0}{3}{10}
        \mysquare{0}{4}{8}
        \mysquare{0}{5}{3}
        \mysquare{0}{6}{1}

        \mysquare{1}{0}{1}
        \mysquare{1}{1}{3}
        \mysquare{1}{2}{8}
        \mysquare{1}{3}{10}
        \mysquare{1}{4}{8}
        \mysquare{1}{5}{3}
        \mysquare{1}{6}{1}


        \mygrid
        \foreach \i in {\yMin,...,\yMaxMin} {
            \draw node [left,black] at (\xMin,\i + 0.5) {$\i$};
        }

        \draw [black,rotate=90] node at (6.5,0.5) {$m \rightarrow$};

        \draw[*->,line width=0.5mm] (0.4,3.5) -- (1.5,3.5);
        \draw[->,line width=0.5mm] (1.5,3.5) -- (1.5,4.5);
        \draw[->,line width=0.5mm] (1.5,4.5) -- (1.5,5.5);
        \draw[->,line width=0.5mm] (1.5,3.5) -- (1.5,2.5);
        \draw[->,line width=0.5mm] (1.5,2.5) -- (1.5,1.5);

        \draw [black] node at (1,-0.5) {\large a)};
    \end{tikzpicture}
    \begin{tikzpicture}[scale=\myscale, every node/.style={transform shape}]

        \mysquare{0}{0}{3}
        \mysquare{0}{1}{8}
        \mysquare{0}{2}{10}
        \mysquare{0}{3}{8}
        \mysquare{0}{4}{3}
        \mysquare{0}{5}{1}
        \mysquare{0}{6}{1}

        \mysquare{1}{0}{1}
        \mysquare{1}{1}{3}
        \mysquare{1}{2}{8}
        \mysquare{1}{3}{10}
        \mysquare{1}{4}{8}
        \mysquare{1}{5}{3}
        \mysquare{1}{6}{1}


        \mygrid

        \draw[*->,line width=0.5mm] (0.4,2.5) -- (1.5,2.5);
        \draw[->,line width=0.5mm] (0.5,1.5) -- (1.5,1.5);
        \draw[->,line width=0.5mm] (0.5,3.5) -- (1.5,3.5);
        \draw[->,line width=0.5mm] (1.5,3.5) -- (1.5,4.5);
        \draw[->,line width=0.5mm] (1.5,4.5) -- (1.5,5.5);

        \draw [black] node at (1,-0.5) {\large b)};
    \end{tikzpicture}
    \begin{tikzpicture}[scale=\myscale, every node/.style={transform shape}]

        \mysquare{0}{0}{5}
        \mysquare{0}{1}{10}
        \mysquare{0}{2}{5}
        \mysquare{0}{3}{1}
        \mysquare{0}{4}{4}
        \mysquare{0}{5}{8}
        \mysquare{0}{6}{4}

        \mysquare{1}{0}{5}
        \mysquare{1}{1}{10}
        \mysquare{1}{2}{5}
        \mysquare{1}{3}{1}
        \mysquare{1}{4}{4}
        \mysquare{1}{5}{8}
        \mysquare{1}{6}{4}


        \mygrid

        \draw[*->,line width=0.5mm] (0.4,1.5) -- (1.5,1.5);
        \draw[->,line width=0.5mm] (0.5,5.5) -- (1.5,5.5);
        \draw[->,line width=0.5mm] (1.5,1.5) -- (1.5,2.5);
        \draw[->,line width=0.5mm] (1.5,1.5) -- (1.5,0.5);
        \draw[->,line width=0.5mm] (1.5,5.5) -- (1.5,4.5);
        \draw[->,line width=0.5mm] (1.5,5.5) -- (1.5,6.5);

        \draw [black] node at (1,-0.5) {\large c)};
    \end{tikzpicture}
    \begin{tikzpicture}[scale=\myscale, every node/.style={transform shape}]

        \mysquare{0}{0}{3}
        \mysquare{0}{1}{3}
        \mysquare{0}{2}{3}
        \mysquare{0}{3}{3}
        \mysquare{0}{4}{3}
        \mysquare{0}{5}{3}
        \mysquare{0}{6}{3}

        \mysquare{1}{0}{8}
        \mysquare{1}{1}{8}
        \mysquare{1}{2}{8}
        \mysquare{1}{3}{8}
        \mysquare{1}{4}{8}
        \mysquare{1}{5}{8}
        \mysquare{1}{6}{8}


        \mygrid

        \draw[*->,line width=0.5mm] (0.4,0.5) -- (1.5,0.5);
        \draw[->,line width=0.5mm] (1.5,0.5) -- (1.5,1.5);
        \draw[->,line width=0.5mm] (1.5,1.5) -- (1.5,2.5);
        \draw[->,line width=0.5mm] (1.5,2.5) -- (1.5,3.5);
        \draw[->,line width=0.5mm] (1.5,3.5) -- (1.5,4.5);
        \draw[->,line width=0.5mm] (1.5,4.5) -- (1.5,5.5);
        \draw[->,line width=0.5mm] (1.5,5.5) -- (1.5,6.5);

        \draw [black] node at (1,-0.5) {\large d)};
    \end{tikzpicture}
    \begin{tikzpicture}[scale=\myscale, every node/.style={transform shape}]

        \mysquare{0}{0}{8}
        \mysquare{0}{1}{8}
        \mysquare{0}{2}{8}
        \mysquare{0}{3}{8}
        \mysquare{0}{4}{8}
        \mysquare{0}{5}{8}
        \mysquare{0}{6}{8}

        \mysquare{1}{0}{3}
        \mysquare{1}{1}{3}
        \mysquare{1}{2}{3}
        \mysquare{1}{3}{3}
        \mysquare{1}{4}{3}
        \mysquare{1}{5}{3}
        \mysquare{1}{6}{3}


        \mygrid

        \draw[*->,line width=0.5mm] (0.4,0.5) -- (1.5,0.5);
        \draw[->,line width=0.5mm] (0.4,1.5) -- (1.5,1.5);
        \draw[->,line width=0.5mm] (0.4,2.5) -- (1.5,2.5);
        \draw[->,line width=0.5mm] (0.4,3.5) -- (1.5,3.5);
        \draw[->,line width=0.5mm] (0.4,4.5) -- (1.5,4.5);
        \draw[->,line width=0.5mm] (0.5,5.5) -- (1.5,5.5);
        \draw[->,line width=0.5mm] (0.4,6.5) -- (1.5,6.5);

        \draw [black] node at (1,-0.5) {\large e)};
    \end{tikzpicture}